\documentclass[conference]{IEEEtran}
\IEEEoverridecommandlockouts

\usepackage{cite}
\usepackage{amsmath,amssymb,amsfonts}
\usepackage{algorithmic}
\usepackage{graphicx}
\usepackage{textcomp}

\usepackage{subcaption,soul}
\usepackage{tikz}

\def\BibTeX{{\rm B\kern-.05em{\sc i\kern-.025em b}\kern-.08em
T\kern-.1667em\lower.7ex\hbox{E}\kern-.125emX}}

\begin{document}

\title{Can Quantum Receiver Beat the SIC Limit in Multiple Access Networks?}

\author{
\IEEEauthorblockN{Shiqian Guo}
\IEEEauthorblockA{
\textit{Department of Computer Science} \\
\textit{North Carolina State University}\\
Raleigh, NC, USA \\
sguo26@ncsu.edu
}
\and
\IEEEauthorblockN{Huaiyu Dai}
\IEEEauthorblockA{
\textit{Department of Electrical and Computer Engineering} \\
\textit{North Carolina State University}\\
Raleigh, NC, USA \\
hdai@ncsu.edu
}
\and
\IEEEauthorblockN{Jianqing Liu}
\IEEEauthorblockA{
\textit{Department of Computer Science} \\
\textit{North Carolina State University}\\
Raleigh, NC, USA \\
jliu96@ncsu.edu
}
}

\maketitle

\begin{abstract}
Successive interference cancellation (SIC) is an important technique for 5G/B5G wireless receivers to resolve interfering signals from multiple users. While SIC has been proven to approach the channel capacity limit in many settings of multiple access networks, it remains unknown if this limit can be surpassed by an advanced yet practically implementable quantum receiver technique. In this work, we answer this quest by proposing a full-stack quantum receiver that integrates front-end quantum sensing with back-end quantum signal processing. The proposed technique leverages simple qubit ensembles without using any complex entanglement resources to parallelly extract interfering multiuser signals, with a channel capacity beyond the limit of SIC in some operational corners. In performance evaluation, we report notable quantum advantage over the ultimate multiple-access channel capacity limit and practical SIC algorithms in low-SNR regimes in terms of spectral efficiency and detection efficiency due to the receiver's ability to exploit quantum correlated measurements and superposition-enabled parallel processing.
\end{abstract}

\maketitle

\section{Introduction}
SIC is an important receiver technique, which was first suggested by Thomas Cover in \cite{cover2003broadcast} and has thus far been shown to achieve the capacity limit of several classes of multiuser channels. The key idea of SIC is that users are decoded sequentially, with the receiver canceling interference after each user. For example, it may detect the strongest signal first, treat others as noise, subtract the decoded signal from the received signal, and repeat the process until all users' signals are detected. A perfect SIC hinges on several ideal conditions, such as distinct received power levels, capacity-achieving encoding, and accurate decoding at each stage. In modern wireless networks, an SIC receiver can be combined with many multiple access technologies such as orthogonal frequency division multiplexing (OFDM) and non-orthogonal multiple access (NOMA) techniques. 

Although SIC can approach the multiple-access channel capacity under strict assumptions, this capacity bound implicitly assumes classical Gaussian measurements whose noise floor is bounded by the Standard Quantum Limit (SQL). In past studies, it has been recognized that a quantum receiver that exploits non-Gaussian measurement can beat SQL and achieve the Holevo capacity, exceeding the classical multiple-access channel capacity bound. The intuition is straightforward. Multiple-access channel capacity assumes a fixed measurement process whose resulting classical signal is corrupted by Gaussian noise that captures physical imperfections prior to or during measurement, while Holevo capacity accounts for all possible measurements and limits the information extractable from the quantum state before measurement. Although it is promising in theory, it remains unknown how to beat a perfect SIC receiver (or multiple-access channel capacity), if not fully attain the ultimate Holevo capacity, using a practically feasible quantum receiver. The key to this research problem is to develop an efficient protocol to perform non-Gaussian measurement, thus overcoming the classical i.i.d. Gaussian noises in independent measurements. Existing methods are mainly based on quantum entanglement, a non-classical resource that can result in correlated measurements at the receiver. However, quantum entanglement is a complex and fragile resource to create and maintain coherent. An alternative entanglement-free technique is to create temporally correlated measurements using adaptive feedback control \cite{guo2026quantum,guo2025two}, but the control overhead is daunting. A natural research question to ask is if a noisy intermediate-scale quantum  receiver can beat the SIC capacity limit without using complex quantum resources like entanglement. This paper attempts to answer this question for the first time in the community.


Like a classical wireless receiver, a quantum receiver consists of front-end signal detecting and back-end signal processing units. For the front-end unit, quantum sensing based on electromagnetic-sensitive qubits like Rydberg atoms is capable of detecting weak wireless signals with sensitivity beyond the classical dipole antenna \cite{degen2017quantum}. However, solely leveraging the high-sensitivity characteristics of quantum sensing does not result in quantum advantage over the SIC receiver, because power interference will be detected with the same level of sensitivity and cannot be stripped off. Therefore, the back-end unit must collaborate to dismantle the intertwined multiuser signals. In light of it, to build a synergy between the front-end and back-end units, we propose a quantum receiver architecture that combines quantum sensing with quantum signal processing. The most salient characteristics of this quantum receiver is that after the sensing qubits capture the intertwined multiuser signals from the air, the state of the sensing qubits is not measured into classical information but rather continuing to propagate to the back-end quantum circuits for processing. In so doing, we reduce measurement noise and information loss arising from the state collapse of the sensing qubits during measurement. This will preserve as much quantum information till the very end to attain quantum advantage. 

\section{State of the Art}
\subsection{SIC Algorithms and Limits}
The theoretical performance of SIC has been extensively studied~\cite{Cover2006}, but in practice it is limited by noise accumulation and error propagation \cite{Goldsmith2005, Ding2016impact}. At low SNR, unreliable early-stage decoding leads to imperfect interference cancellation and residual interference that degrades subsequent detections. To address these issues, more advanced successive and joint detection methods have been proposed. In particular, joint maximum likelihood detection (JMLD) achieves optimal performance by jointly decoding all users under ideal conditions, but its exponential complexity in both user number and modulation order makes it impractical for large-scale systems.

To balance optimality and complexity, structured detection methods have been developed. Sphere decoding achieves near-ML performance with reduced complexity in moderate dimensions and high-SNR regimes \cite{Viterbo1999, Hassibi2005}, while lattice-reduction-aided detection improves linear receivers through channel orthogonalization, approaching ML performance with polynomial complexity \cite{Yao2002, Windpassinger2004}. In parallel, enhanced SIC variants, including soft, iterative, and MMSE-based SIC, reduce error propagation through reliability-aware refinement \cite{Ding2016}. Message-passing detectors exploit probabilistic graphical models for near-capacity multiuser detection \cite{Kschischang2001, Fletcher2018}, while learning-based methods use neural networks to approximate optimal detection and interference cancellation under channel uncertainty \cite{Samuel2017, Nachmani2018}.

\subsection{Quantum Signal Processing}
In quantum signal processing (QSP), estimating unknown quantum dynamics is important for wireless communication problems, where techniques such as quantum principal component analysis (QPCA) extract dominant eigen-states and eigenvalues of unknown density matrices for spectral decomposition \cite{lloyd2014quantum, he2022low}. However, QPCA requires multiple state copies and controlled-SWAP operations, making it noise-sensitive and impractical in weak-signal regimes. Quantum phase estimation (QPE) improves spectral resolution using filtering methods like Gaussian smoothing, but still relies on deep coherent circuits and repeated controlled unitaries, which limits near-term applicability \cite{ding2024quantum}. More recent QSP-based and measurement-driven approaches reduce circuit depth by using polynomial transformations or direct statistical reconstruction (e.g., classical shadows), improving robustness and practicality in noisy quantum hardware \cite{dong2025optimal, shen2025estimating, shen2026efficient}.


\subsection{Integrated Quantum Sensing and Processing}

Recent progress in quantum technologies has highlighted the benefits of tightly integrating front-end quantum sensing with back-end quantum processing to enhance overall system performance. In particular, quantum  sensing frameworks leverage quantum resources not only to improve field sensitivity but also to enable downstream information extraction through quantum or quantum-inspired processing pipelines. For example, Allen \textit{et al.} \cite{allen2025quantum} provide a broad overview of QSP–enhanced sensing architectures, emphasizing how quantum processors can be used to post-process measurement data and extract weak signal features beyond classical limits. 

Beyond general architectures, recent works have explored more structured quantum information processing strategies for multiparameter estimation. Gong \textit{et al.} \cite{gong2026robust} use quantum scrambling to improve robust multiparameter estimation under noise, while Mokeev \textit{et al.} \cite{mokeev2026enhancing} incorporate quantum computation into optical imaging pipelines. Together, these studies highlight unified quantum sensing–processing frameworks with jointly designed measurement and inference.

\subsection{Research Gap and Scope of this Work}
\begin{figure*}[htpb]
	\centering
		\includegraphics[width=\linewidth]{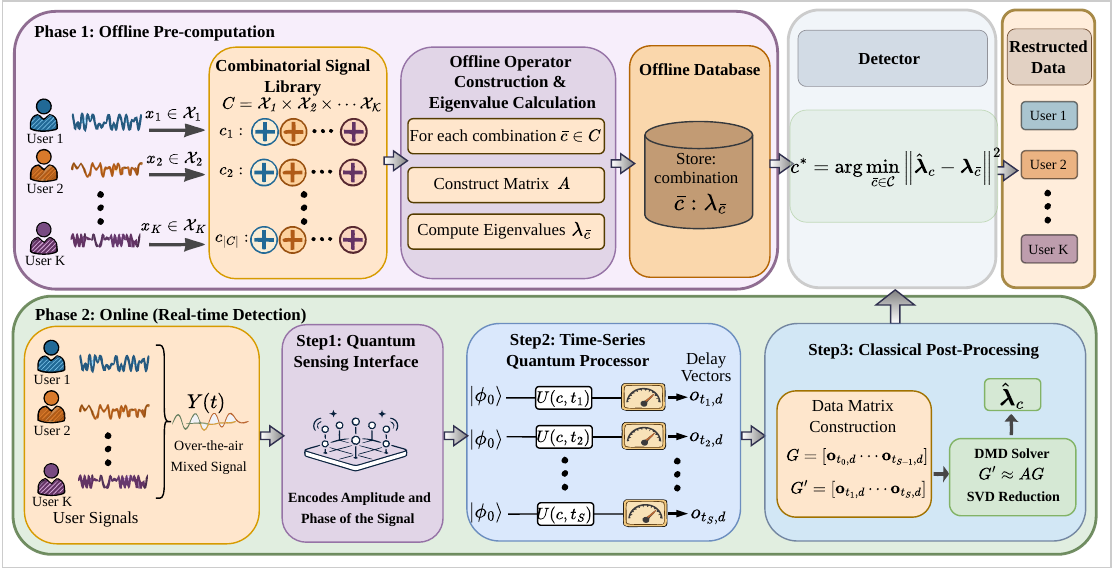}
	\caption{Working principle of the proposed quantum receiver to detect interfering multiuser signals.}
    \label{fig: multiuser comm system}
\end{figure*} 
Recent advances in integrated quantum sensing and quantum processing have demonstrated promising directions for enhancing measurement and inference capabilities.
However, these existing frameworks primarily focus on single-source sensing, imaging, or parameter estimation tasks, where the goal is to solely enhance the accuracy of a single quantum or physical field measurement. They do not explicitly address the challenges arising in multi-source discrimination scenarios, with the wireless multiuser non-orthogonal access being a notable example where multiple signals are deeply mixed and must be separated under the constraints of noise, interference, and limited measurement resources. In particular, current integrated quantum sensing–processing architectures lack a principled mechanism for structured multi-access signal separation and decoding in a networking context.

In this work, we bridge this gap by proposing a quantum receiver architecture for wireless multiple access networks that integrates front-end quantum sensing with back-end quantum signal processing. The front-end quantum sensing layer perceives the incoming multiuser signal as a measurable quantum dynamics, enabling phase- and amplitude-sensitive encoding in the sensing qubits. The back-end quantum signal processing module applies parallel Hadamard tests and dynamic mode decomposition (DMD) to perform structured projection and combinatorial hypothesis testing for user separation and decoding, effectively enabling a quantum-enhanced successive interference handling strategy.

Unlike prior work of single sensing or signal processing objectives such as parameter estimation or image reconstruction, our proposed quantum receiver aims to solve a complex networking problem, where the objective is to resolve interfering signals in an end-to-end manner. This establishes a new direction for integrated quantum sensing–processing systems promised to achieve quantum advantage in wireless multiple access networks, further opening doors to many engineering problems in the next-generation wireless systems.

\section{Principles of Quantum Receiver}\label{sec:system}
\subsection{System Overview} 
Our proposed quantum receiver architecture is schematically illustrated in Fig.~\ref{fig: multiuser comm system}. This architecture consists of two phases: offline dictionary buildup and online inference. The offline phase is a purely classical process in which different symbol combinations from multiple users and their eigenvalues from a special matrix (this will become clear in Section III.C.3) are pre-computed and stored. In contrast, the online phase describes how the proposed quantum receiver works. Basically, it consists of three technical steps. First, instead of using entanglement in front-end quantum sensing, we propose an ensemble of qubits that are spatially separated. This is analogous to an antenna array in a classical receiver. The over-the-air multiuser signal, also known as the system Hamiltonian in the quantum physics terminology, will drive each qubit to a unique quantum state so that it embeds the phase and amplitude of the multiuser signal. Second, we propose a quantum circuit inspired by the Hadamard test to extract the real and imaginary parts of the quantum states associated with the sensing qubits.  Third, the extracted information is processed by a dynamic mode
decomposition algorithm, whose role is to perform eigenvalue decomposition for measurement data from the Hadamard test in Step 2. The set of extracted eigenvalues is implicitly correlated with the set of user signals. Therefore, at the end of the eigenvalue analysis, a maximum likelihood detector (or any classical classifier) can be adopted to compare with the offline-prepared dictionary for signal detection. 

As illustrated in Fig.~\ref{fig: multiuser comm system}, we consider an uplink multiple-access communication scenario in which $K$ users share the same communication resources to transmit their individual signals $x_k(t)$. The transmitted waveform for the $k$-th user is generated based on a discrete amplitude-phase modulation scheme and is given by $x_k(t) = \alpha_k \cos(\omega t + \theta_k)$,
where the amplitude $\alpha_k$ and phase $\theta_k$ are selected from predefined discrete sets according to the transmitted information symbol. Here, $\omega$ denotes the angular frequency of the carrier signal.

Let $x_k \in \mathcal{X}_k$ denote the discrete information symbol of the $k$-th user drawn from a finite modulation alphabet $\mathcal{X}_k$. Each symbol $x_k$ uniquely determines the waveform parameters $(\alpha_k, \theta_k)$, and thus fully specifies the transmitted signal $x_k(t)$.

For convenience, we adopt a symbol-level representation and use $x_k$ to denote the transmitted symbol in the following derivations. Let $c = [x_1, x_2, \ldots, x_K]$
denote the transmitted symbol vector of all users, and let $c^{*} = [\hat{x}_1, \hat{x}_2, \ldots, \hat{x}_K]$
denote the estimated symbol vector at the quantum (or classical) receiver.

\subsection{Classical SIC Receiver}\subsubsection{Noise model}
In classical wireless receivers, the received signal $y(t)$ is the addition of all users’ transmitted signals over a shared channel, corrupted by AWGN, i.e., 

\begin{equation}
y(t) = \sum_{k=1}^{K} \sqrt{P_k} h_k \cdot x_k(t) + n(t),
\end{equation}
where $P_k$ is the transmit power scaling factor, $h_k$ is the channel coefficient, AWGN is modeled as $n(t) \sim \mathcal{CN}(0, \sigma_n^2)$, where $\sigma_n^2$ is normalized to unity without loss of generality.
The channel is modeled as a distance-dependent path-loss channel, where the channel coefficient of the $k$-th user is given by $h_k = l_k^{-\xi/2}$,
where $l_k$ denotes the distance between the $k$-th user and the receiver, and $\xi$ is the path-loss exponent. For simplicity, other small-scale channel fading like phase variations is neglected and left for future investigations.

\subsubsection{SIC algorithms}
For an SIC receiver, users are decoded sequentially according to a predefined order. At each step, the previously decoded signals are subtracted from the received signal. Specifically, the $k$-th user symbol is detected as
\begin{equation}
\hat{x}_k = \arg \min_{x_k \in \mathcal{X}_k}
\left|
h_k^{H} y - h_k^{H} \sum_{i=1}^{k-1} \sqrt{P_i} h_i \hat{x}_i - \sqrt{P_k} \|h_k\|^2 x_k
\right|^2,
\end{equation}
where $\hat{x}_i$ denotes the previously detected symbols. After detecting $\hat{x}_k$, its contribution $\sqrt{P_k} h_k \hat{x}_k$ is subtracted from the received signal before decoding the next user.

A variation of SIC is parallel interference cancellation (PIC), which performs simultaneous multiuser detection by iteratively estimating and canceling multiuser interference from all users. Unlike optimal JMLD, PIC is generally a low-complexity suboptimal detector. However, JMLD provides the optimal benchmark, and solves the following problem:
\begin{equation}
\{\hat{x}_1, \hat{x}_2, \ldots, \hat{x}_K\}
= \arg \min_{x_k \in \mathcal{X}_k}
\left\|
y - \sum_{k=1}^{K} \sqrt{P_k} h_k x_k
\right\|^2.
\end{equation}

\subsection{Proposed Quantum Receiver}
\subsubsection{Noise model}
The front-end quantum sensor $s$ receives a mixed signal $c$ from $K$ users, which collectively drives the quantum state of each sensing qubit for up to a time duration of $t_s = s\Delta t$. $\Delta t$ is the time sampling interval. The rationale for selecting a sensing duration of $t_s$ will become clear in subsequent sections. We can then calculate the net positive/negative bias of the mixed signal for a sensing time $t_s$ as follows
\begin{equation}\label{eq:net-bias}
\begin{aligned}
\rho(c,t_s) 
&= \int_{0}^{s \Delta t} \sum_{k=1}^{K} \sqrt{P_k}\, h_k \alpha_k 
\cos(\omega t + \theta_k)\, dt \\
&= \sum_{k=1}^{K} \frac{\alpha_k}{\omega}
\left[ \sin(\omega s \Delta t + \theta_k) - \sin(\theta_k) \right].
\end{aligned}
\end{equation}
Using an electronic spin qubit as an example, if a positive (resp., negative) wireless signal drives the spin to rotate clockwise (resp., counter-clockwise), the net bias over a time period $[0, t_s]$ in Eq.~(\ref{eq:net-bias}) defines the effective signal strength that can result in a net spin angle after the sensing time.

Under the external field in Eq.~(\ref{eq:net-bias}), we adopt the widely used one-dimensional transverse-field Ising model (TFIM) to describe the evolution of the quantum state of the sensing qubits \cite{shen2026efficient}. For the front-end quantum sensor of \(N\) sensing qubits under open boundary conditions, the system Hamiltonian is defined as:

\begin{equation}\label{eq:hamiltonian}
H(c,t_s) = -J\sum_{n=1}^{N-1} Z_n Z_{n+1} - \rho(c,t_s) \sum_{n=1}^{N} X_n,
\end{equation}
where \(J\) is the coupling strength between neighboring qubits. Because electromagnetic-sensitive qubits are typically charged particles such as ions and spins, the coupling is caused by the Coulomb force between nearby qubits. The term \(Z_n Z_{n+1}\) represents the interaction between two adjacent qubits \(n\) and \(n+1\), where \(Z_n\) is the Pauli-\(Z\) operator acting on qubit \(n\). This coupling is one-dimensional when qubits are placed in a chain. Effectively, it captures the cross-talk noise between adjacent qubits, which forces the system out of a predefined basis (e.g., X basis) and causes measurement uncertainty. 
Besides, the signal of interest is \(\rho(c, t_s)\) that controls the global transverse field. The operator \(X_n\) is the Pauli-\(X\) operator acting on qubit \(n\). Thus, the signal drives all qubits along the \(X\)-direction.

It is worth mentioning that the TFIM dynamics can be applied to several quantum platforms, such as trapped-ion systems, NV center systems, superconducting qubits, Rydberg atom arrays, and cold-atom optical lattices.


\subsubsection{Front-end quantum sensing}
\begin{figure}[htpb]
	\centering
    \includegraphics[width=\linewidth]{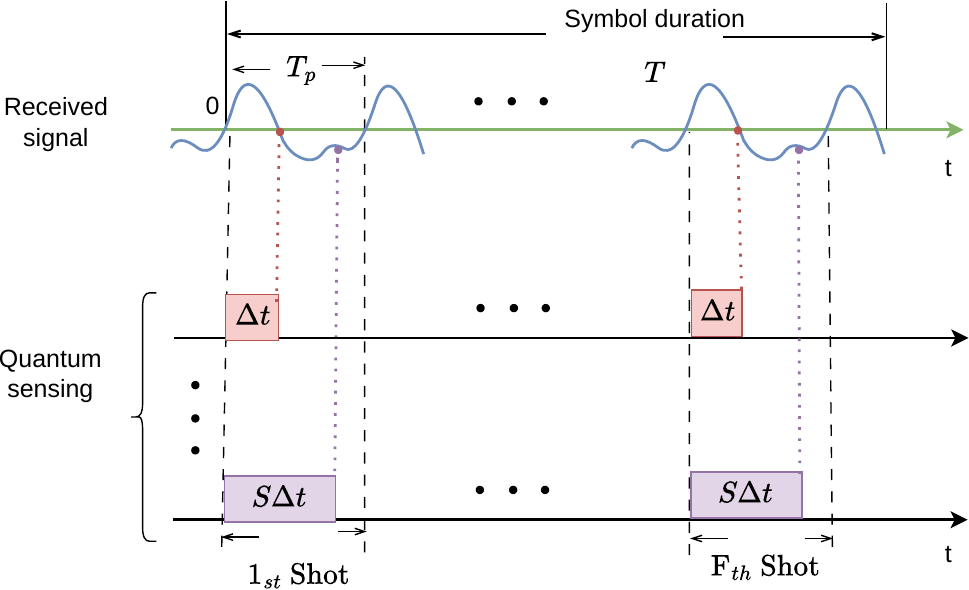}
	\caption{Sensing qubits' interrogation pattern.}
    \label{fig: Timeline QS}
\end{figure} 
 
In the proposed front-end quantum sensing unit, the received mixed signal is mapped onto an effective Hamiltonian in Eq.~(\ref{eq:hamiltonian}) that governs the evolution of the sensing qubit system. Specifically, signal parameters including amplitude, phase, and frequency are encoded as time-dependent control terms in the Hamiltonian such that the quantum dynamics are directly driven by the incident field.

To obtain statistically reliable measurements, each sensing qubit performs multiple measurement shots \(F\) within the symbol duration \(T = F T_p\), where \(T_p\) denotes one signal period. Each sensing qubit adopts a distinct sensing duration that is an integer multiple of \(\Delta t\). As illustrated in Fig.~\ref{fig: Timeline QS}, suppose the mixed signal lasts for a duration \(T\) (e.g., \(4\,\mu s\) for an OFDM symbol). We assume that the maximum sensing duration per shot, denoted by \(S\Delta t\), is constrained to be within one signal period \(T_p\).  There are $S$ group of sensing qubits. This design enables the acquisition of temporally diverse samples while maintaining low sensing latency. From a physical perspective, $S\Delta t$ must also remain below the coherence time of the sensing qubit to avoid significant decoherence effects.

Within each measurement shot, the sensing system is initialized in a fully separable superposition state in the computational basis:
\begin{equation}
|\phi_0\rangle = |+\rangle^{\otimes N}
= \frac{1}{2^{N/2}} \sum_{b \in \{0,1\}^N} |b\rangle,
\label{eq:ini state}
\end{equation}
where $|+\rangle = \frac{1}{\sqrt{2}}(|0\rangle + |1\rangle)$. This state is a tensor product of identical single-qubit superpositions and is therefore non-entangled.

Under the signal-induced interaction, the sensing state evolves coherently as
\begin{equation}
|\phi(c,t_s)\rangle = \exp\left(-i H(c,t_s)\right) |\phi_0\rangle.
\end{equation}

During this evolution, the sensing qubits accumulate information about the received signal through phase accumulation, frequency shifts, and energy redistribution governed by the signal-dependent Hamiltonian. After an interaction time $t_s = s\Delta t$, the encoded information is extracted via measurement. The corresponding observable is given by
\begin{equation}
o(c,s\Delta t) = \langle \phi_0 | U(c, s\Delta t) | \phi_0 \rangle, 
\quad s = 0,1,\ldots,S,
\label{eq:obs sample}
\end{equation}
where $U(c, s\Delta t) = \exp\left(-i H(c, s\Delta t)\right)$.
This quantity can be efficiently estimated using the Hadamard test or related interferometric measurement circuits.

Each measurement shot is synchronized with the signal period and repeated over $F$ shots within the total symbol duration $T$, satisfying $T \geq F \cdot T_{p}$.

\subsubsection{Back-end quantum processing}
\begin{figure}[htpb]
	\centering
		\includegraphics[width=\linewidth]{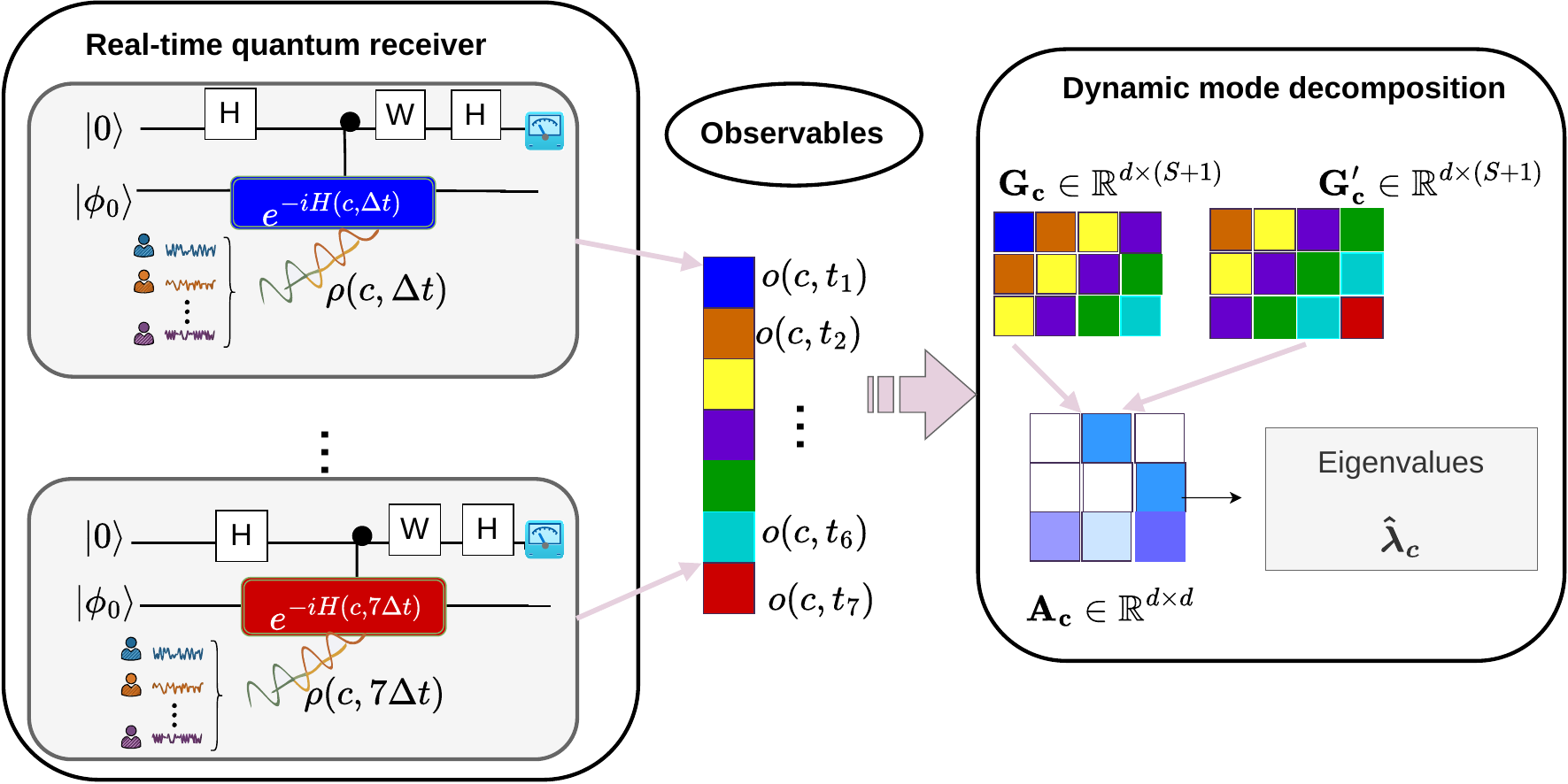}
	\caption{Quantum measurement and processing.}
    \label{fig: Real_time QR_wDMD}
\end{figure} 
The left panel of Fig.~\ref{fig: Real_time QR_wDMD} illustrates the Hadamard-test circuit used to measure the observable $o(c,s\Delta t)$. The joint system is initialized in the product state $|0\rangle_a \otimes |\phi_0\rangle$, where $|0\rangle_a$ denotes the ancilla qubit and $|\phi_0\rangle$ is the initial state of the $N$-qubit sensing register.

A Hadamard gate $\text{H}$ is first applied to the ancilla, generating a superposition state $\text{H}|0\rangle_a = \frac{1}{\sqrt{2}}(|0\rangle_a + |1\rangle_a)$
so that the total state becomes $\frac{1}{\sqrt{2}}\left(|0\rangle_a + |1\rangle_a\right)\otimes|\phi_0\rangle$.

Next, a controlled unitary operation is applied, where the signal-dependent evolution $U(c,s\Delta t)$
acts on the sensing register conditioned on the ancilla being in state $|1\rangle_a$. This produces the entangled state
\[\frac{1}{\sqrt{2}}\left(|0\rangle_a \otimes |\phi_0\rangle + |1\rangle_a \otimes U(c,s\Delta t)|\phi_0\rangle\right).\]

A second Hadamard gate is then applied to the ancilla, mapping the quantum overlap between $|\phi_0\rangle$ and $U(c,t)|\phi_0\rangle$ into measurable population differences, yielding
\[
\frac{1}{2}\Big[|0\rangle_a \otimes (|\phi_0\rangle + U(c,s\Delta t)|\phi_0\rangle)
+ |1\rangle_a \otimes (|\phi_0\rangle - U(c,s\Delta t)|\phi_0\rangle)\Big].
\]

Measurement of the ancilla in the computational ($Z$) basis gives $P(0)=\frac{1}{2}\left(1+\mathrm{Re}\langle \phi_0|U(c,s\Delta t)|\phi_0\rangle\right)$,
from which the observable $o(c,s\Delta t)$ is estimated as $\mathrm{Re}[o(c,t)] = 2P(0)-1$.

For estimating the real part, we set \( W = I \), i.e., the identity gate. To access the imaginary part of \( o(c,t) \), a phase gate \( W = \mathrm{diag}(1,-i) \) is applied to the ancilla prior to the final Hadamard operation, enabling the extraction of \( \mathrm{Im}[o(c,t)] \) using the same measurement procedure.

To characterize the quantum dynamics induced by the true mixed-signal $c$, we apply DMD to the measured observable sequence, which we refer to as observable DMD (ODMD).

As shown in the right panel of Fig.~\ref{fig: Real_time QR_wDMD}, first, the observable $o(c,t)$ is sampled at discrete time instants $\{t_s = s\Delta t\}_{s=0}^{S}$ to form a sequence $\{o(c,t_s)\}$. To capture the underlying system dynamics, a time-delay embedding is constructed by stacking consecutive measurements into delay vectors
\begin{equation}
\mathbf{o}_{t_s,d} =
\begin{bmatrix}
o(c,t_s) \\
o(c,t_{s+1}) \\
\vdots \\
o(c,t_{s+d-1})
\end{bmatrix}, \quad 0 \le s \le S,
\end{equation}
where $d$ is the embedding dimension.

Using these delay vectors, we form two time-shifted data matrices
\[
\begin{aligned}
&G =
\begin{bmatrix}
\mathbf{o}_{t_0,d} & \mathbf{o}_{t_1,d} & \cdots & \mathbf{o}_{t_{S-1},d}
\end{bmatrix}, \\
&G' =
\begin{bmatrix}
\mathbf{o}_{t_1,d} & \mathbf{o}_{t_2,d} & \cdots & \mathbf{o}_{t_{S},d}
\end{bmatrix}.
\end{aligned}
\]

DMD approximates the underlying nonlinear dynamics by identifying a best-fit linear operator $A$ that satisfies $G' \approx A G$, where $A$ is obtained via a least-squares solution
\begin{equation}\label{eq:A_LS}
A = G' G^{+},
\end{equation}
and $(\cdot)^{+}$ denotes the Moore--Penrose pseudoinverse.

The eigenvalues of $A$, $\boldsymbol{\lambda}_c $
capture the dominant dynamical modes of the quantum system and serve as spectral features characterizing the signal-induced evolution. Based on these features, signal detection is performed by matching the estimated eigenvalue vector $\hat{\boldsymbol{\lambda}}_{c}$ with a precomputed dictionary $\{\boldsymbol{\lambda}_{\bar{c}}\}_{\bar{c} \in \mathcal{C}}$, i.e.,
\begin{equation}
{c^{*}} = \arg\min_{\bar{c} \in \mathcal{C}} 
\left\| \hat{\boldsymbol{\lambda}}_{c} - \boldsymbol{\lambda}_{\bar{c}} \right\|^2.
\label{eq:lambda ML}
\end{equation}

To mitigate the effects of both statistical noise and quantum hardware noise, we adopt the least-squares formulation in Eq.~(\ref{eq:A_LS}), which provides a rigorous and robust regularization framework. The recovery of the system matrix $A$ requires computing the pseudo-inverse of the data matrix $G$, whose conditioning critically affects the stability of the solution.

We perform a singular value decomposition (SVD) of $G$:
\begin{equation}
G = \sum_{\ell=0}^{d-1} \sigma_\ell \, \mathbf{u}_\ell \mathbf{v}_\ell^\dagger,
\label{eq:svd}
\end{equation}
where $\sigma_\ell > 0$ are the singular values, and $\mathbf{u}_\ell$ and $\mathbf{v}_\ell$ denote the corresponding left and right singular vectors, respectively.

To enhance noise resilience, we apply a thresholding procedure to suppress perturbative errors. Specifically, we truncate singular values below a prescribed cutoff $\tilde{\delta} \, \sigma_{\max}(G)$, where $\tilde{\delta} > 0$ is a relative threshold parameter. This yields the regularized matrix:
\begin{equation}
G \;\mapsto\; G_{\tilde{\delta}} 
= \sum_{\ell:\,\sigma_\ell > \tilde{\delta}\,\sigma_{\max}} 
\sigma_\ell \, \mathbf{u}_\ell \mathbf{v}_\ell^\dagger,
\label{eq:truncated_svd}
\end{equation}
where $\sigma_{\max} = \max_\ell \sigma_\ell$ denotes the largest singular value of $G$.

\subsection{Computational Complexity}
The computational complexity of the proposed ODMD algorithm consists of two components. First, the ODMD-based feature extraction requires
\begin{equation}
\mathcal{O}(dS\varsigma + d^3)
\end{equation}
operations, where $d$ is the delay embedding dimension, $S$ is the number of time samples, and $\varsigma$ is the truncated numerical rank. Second, the spectral-domain detection stage performs dictionary matching over all possible transmitted symbols, resulting in complexity
\begin{equation}
\mathcal{O}(|\mathcal{C}| d).
\end{equation}

For a multiuser amplitude-phase modulation system with $K$ users and modulation orders $M$, the dictionary size scales as $|\mathcal{C}| = M^K$,
leading to exponential detection complexity
\begin{equation}
\mathcal{O}(M^K d).
\end{equation}
Consequently, the overall complexity is dominated by the combinatorial growth of the signal hypothesis space, while ODMD feature extraction remains polynomial in the system parameters.

\subsection{Numerical Example}\label{sec:example}
The above discussion is mathematically dense. To provide clearer intuition, we introduce a toy example to illustrate the proposed design. Specifically, consider a simple two-user access network ($K=2$), where 4-ary amplitude modulation is employed with symbol set $\alpha \in \{0.5, 1.0, 1.5, 2\}$. The two users transmit the true signal $c = [00, 01]$. The carrier frequency is set to $\omega = 1$, the time sampling interval is $\Delta t = 0.4$, and the delay parameter is $d = 2$. There are $S = 5$ groups of sensing qubits, each of which contains $N = 2$ sensing qubits. The coupling strength is $J=1$. We prepare an initial quantum state $|\phi_0\rangle$ according to Eq.~(\ref{eq:ini state}) as
\[
|\phi_0\rangle = |+\rangle \otimes |+\rangle=  \frac{1}{2}\left(|00\rangle + |01\rangle + |10\rangle + |11\rangle\right)
\]

After the quantum sensing stage and subsequent quantum processor measurement, we obtain the expectation value of the observable defined in Eq.~(\ref{eq:obs sample}). 
In the simulation, the expectation values are not computed exactly but are estimated under a finite-shot measurement model. Specifically, each observable is evaluated as a complex random variable \(\tilde{o}(t)=\mathcal{Q}\!\left( o(t) + \eta \right)\). The noise term $\eta = \eta_{\mathrm{Re}} + i\,\eta_{\mathrm{Im}}$ models statistical fluctuations arising from a finite number of measurement shots $F$, where $F = 10^4$. Its real and imaginary components are assumed to be independent Gaussian random variables,
\[
\eta_{\mathrm{Re}},\, \eta_{\mathrm{Im}} \sim \mathcal{N}\!\left(0,\; \frac{1 - |o(t)|^2}{2F}\right),
\]
which reflects the quantum projection noise with variance
\begin{equation}
 \mathrm{Var}(\tilde{o}(t)) = \frac{1 - |o(t)|^2}{F}.\label{eq:var obs}   
\end{equation}

Furthermore, to enforce the physical constraint that expectation values lie within the unit disk, we apply an element-wise magnitude clipping operation \(\mathcal{Q}(\cdot)\), defined as
\[
\mathcal{Q}(z) =
\begin{cases}
z, & |z| \leq 1, \\
\dfrac{z}{|z|}, & |z| > 1.
\end{cases}
\]

Subsequently, we generate $G$ and $G'$ according to Eq.~(\ref{eq:G generate}) and Eq.~(\ref{eq:G' generate}),

\begin{equation}
G =
\begin{bmatrix}
\tilde{o}(t_0) & \tilde{o}(t_1) & \tilde{o}(t_2) & \tilde{o}(t_3) \\
\tilde{o}(t_1) & \tilde{o}(t_2) & \tilde{o}(t_3) & \tilde{o}(t_4)
\end{bmatrix}
\label{eq:G generate}
\end{equation}

\begin{equation}
G' =
\begin{bmatrix}
\tilde{o}(t_1) & \tilde{o}(t_2) & \tilde{o}(t_3) & \tilde{o}(t_4) \\
\tilde{o}(t_2) & \tilde{o}(t_3) & \tilde{o}(t_4) & \tilde{o}(t_5)
\end{bmatrix}
\label{eq:G' generate}
\end{equation}

Next, we compute $A$ using (\ref{eq:A_LS}). Finally, we obtain the eigenvalues of $A$. The first two (dominant) eigenvalues of $A$ are listed below.

\[
\begin{aligned}
\Re(\hat{\boldsymbol{\lambda}}_c) &= (-1.27816,\; 2.00263), \\
\Im(\hat{\boldsymbol{\lambda}}_c) &= (-0.71022,\; -1.04738)
\end{aligned}
\]

In the offline phase, we pre-compute all possible signal combinations and store their corresponding eigenvalues in Table~\ref{table:offline eig real} and Table~\ref{table:offline eig img}. Note that during this offline stage, the generated matrices $G$ and $G'$ are noise-free. Finally, the estimated eigenvalues $\hat{\boldsymbol{\lambda}}{c}$ are compared against a precomputed dictionary containing all candidate signal combinations. The transmitted signal is then determined via maximum-likelihood matching as defined in Eq.~(\ref{eq:lambda ML}).
\begin{table}[t]
\centering
\footnotesize
\setlength{\tabcolsep}{3pt}
\renewcommand{\arraystretch}{1.1}
\begin{tabular}{c|cccc}
 & 00 & 01 & 10 & 11 \\
\hline
00 &
$0.353,\;0.638$ &
$-1.303,\;1.994$ &
$1.445,\;1.676$ &
$0.094,\;1.360$ \\

01 &
\hl{$-1.303,\;1.994$} &
$1.445,\;1.676$ &
$0.094,\;1.360$ &
$-1.611,\;1.636$ \\

10 &
$1.445,\;1.676$ &
$0.094,\;1.360$ &
$-1.611,\;1.636$ &
$-2.053,\;0.852$ \\

11 &
$0.094,\;1.360$ &
$-1.611,\;1.636$ &
$-2.053,\;0.852$ &
$-0.911,\;0.677$ \\
\end{tabular}
\caption{Real parts of eigenvalues.}
\label{table:offline eig real}
\end{table}

\begin{table}[t]
\centering
\footnotesize
\setlength{\tabcolsep}{3pt}
\renewcommand{\arraystretch}{1.1}
\begin{tabular}{c|cccc}
 & 00 & 01 & 10 & 11 \\
\hline
00 &
$2.572,\;-2.329$ &
$-0.645,\;-1.050$ &
$1.351,\;-1.411$ &
$2.047,\;-0.789$ \\

01 &
\hl{$-0.645,\;-1.050$} &
$1.351,\;-1.411$ &
$2.047,\;-0.789$ &
$1.080,\;0.415$ \\

10 &
$1.351,\;-1.411$ &
$2.047,\;-0.789$ &
$1.080,\;0.415$ &
$-0.468,\;1.079$ \\

11 &
$2.047,\;-0.789$ &
$1.080,\;0.415$ &
$-0.468,\;1.079$ &
$-1.770,\;1.352$ \\
\end{tabular}
\caption{Imaginary parts of eigenvalues.}
\label{table:offline eig img}
\end{table}

\section{Performance Evaluation and Discussion}\label{sec:results}
\subsection{Setups and Metrics}
\subsubsection{System Parameters}
Here, we present simulation setups to evaluate the proposed quantum receiver for multiuser communications. The channel gain for each user is set to $h_k = 1$. All users are assumed to share an identical transmit power scaling factor $P_k$. Unless otherwise specified, $P_k = 0.5^2$ is adopted in the following results. The carrier frequency is normalized to $\omega=1$ for simplicity. The transmitted signal employs amplitude-phase modulation with uniformly spaced amplitude and phase levels. The amplitude levels are
\begin{equation*}
\alpha_i = 0.5 + \frac{1.5*(i-1)}{M_{\alpha} - 1} , \quad i = 1,2,\dots,M_{\alpha},
\end{equation*}
where $M_{\alpha}$ denotes the number of amplitude levels. The phase levels are
\begin{equation*}
\theta_p = \frac{p \pi}{M_p}, \quad p = 0,1,\dots,M_p-1.
\end{equation*}
This forms an $M_{\alpha} \times M_p$ amplitude-phase constellation, where each symbol is represented by $(\alpha_i, \theta_p)$.
$M_{\alpha}$ and $M_{p}$ are both set to 4.

The coupling strength $J$ in the TFIM Hamiltonian model is a hardware-specific parameter in the range of [0,1]. In the subsequent evaluation, we will examine how to select relevant hyper-parameters according to $J$, namely \( S \) groups (or, modes hereafter) of sensing qubits, \( N \) sensing qubits assigned to each mode, and measurement shots $F$. Note that sensing qubits across different groups are independent.  The sensing duration for the \( s \)-th mode is given by \( s \cdot \Delta t = s\cdot 0.4 \). Each sensing qubit is coupled with an ancilla qubit for backend quantum processing.

In the ODMD stage, the first three eigenvalues are selected to capture the dominant system dynamics. The data matrices are constructed using a time-delay embedding dimension \( d = \lfloor S / 2.5 \rfloor \), and pairwise thresholding is applied with \( \tilde{\delta} = 10^{-2} \). 

To demonstrate the quantum advantage, we define the classical SIC decoding method and its achievable bound. 
For a fair comparison, the classical detector operates on \( S \cdot N >2 \) time snapshots samples, matching the total number of signal time samples within one signal period used in the quantum system. The sampling interval is \( \Delta t \), leading to a maximum sampling time of \( S \cdot  N  \cdot \Delta t \).

\subsubsection{Classical Baselines} For a fair comparison between quantum and classical receivers, we ensure that the total energy budget per transmitted symbol remains identical across all schemes. Since the classical receiver forms one effective decision statistic per symbol duration, the transmit power $P$ is effectively accumulated over the full symbol duration, giving an effective scaling of $F \cdot P$. Accordingly, the capacity of the multiple access channel\cite{dai2018survey,cover2006elements} is given by
\begin{equation}
V = \log_2 \left(1 + \frac{K \cdot F \cdot P \cdot |x|^2}{n} \right),
\end{equation}
where $x$ denotes the maximum amplitude of transmitted symbol, and $n$ represents the noise power.
Thus, the theoretical SIC bound is expressed as $\min\{V, K \log_2 M\}$, where $K \log_2 M$ represents the maximum achievable information content constrained by the modulation order $M = M_{\alpha} \times M_p$ and the number of users $K$.

\subsubsection{Figures of Merit} The probability of error corresponds to the block error rate, defined as the probability that the entire multiuser symbol vector is incorrectly decoded:
\begin{equation*}
P_{e} = \Pr\{\mathbf{c^{*}} \neq \mathbf{c}\}.
\end{equation*}
Then, we can calculate the network-level spectral efficiency, given by
\[
K(1-P_e)\log_2 M.
\]

The reported results in the following sections are based on averaging the results over 200 independent Monte Carlo simulations.

\subsection{Results and Discussion}
\subsubsection{Hyper-parameter Selection}

\begin{figure}[!t]
\centering
  \includegraphics[width=.5\textwidth]{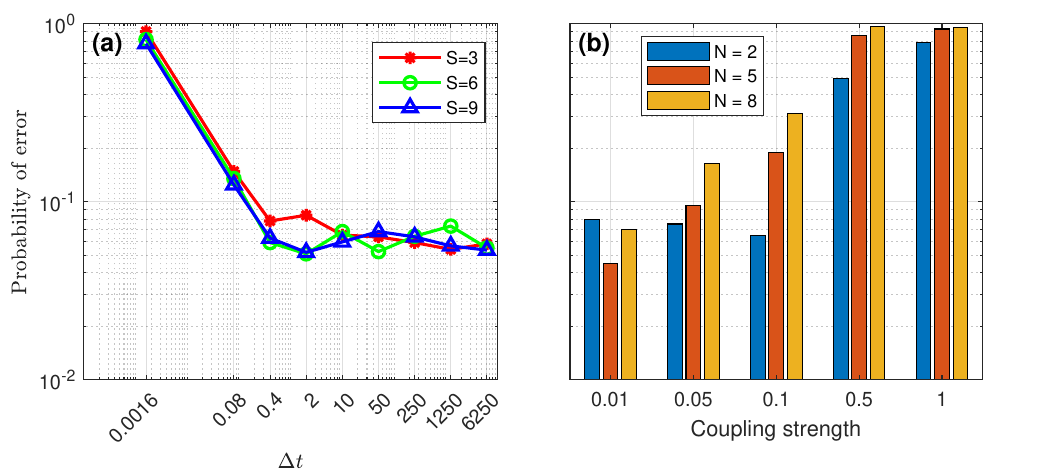}
    \caption{$P_e$ under different sensing modes and rates.}
\label{fig:hyperparameter_selection}
\end{figure}

The results in this section are obtained by setting $K=2$. Fig.~\ref{fig:hyperparameter_selection} illustrates the hyper-parameter selection for the proposed quantum receiver. 
Fig.~\ref{fig:hyperparameter_selection} (a) illustrates the probability of error $P_{e}$ as a function of the time sampling gap $\Delta t$ under different $S$. It is observed that $P_{e}$ decreases as the maximum time span of samples increases (i.e., by increasing either $\Delta t$ or $S$), particularly when the total sampling duration satisfies $S\cdot\Delta t \ll T_{p}$. We set the signal period \(T_p = 2\pi\) without loss of generality. However, when the total sampling duration $S\cdot\Delta t$ approaches or exceeds $T_{p}$, further increasing the sampling span no longer improves $P_{e}$. Moreover, a larger $S\cdot\Delta t$ requires a longer symbol duration $T$ under the same number of measurement shots (or equivalently fewer measurement shots under the same symbol duration), which reduces effective data rate.
Therefore, considering both qubit resources and effective data rate, we select $S=3$ and $\Delta t=0.4$. This setup also implies that the proposed quantum receiver is exempt from sampling over the entire signal period, i.e., $S\cdot\Delta t < T_{p}$, which is otherwise infeasible in classical receivers.

Fig.~\ref{fig:hyperparameter_selection}(b) illustrates the probability of error under different coupling strengths and numbers of sensing qubits. It can be observed that $P_{e}$ increases as the coupling strength $J$ increases, regardless of the number of sensing qubits $N$. This is because $J$ characterizes the crosstalk interaction between adjacent qubits; a larger $J$ drives the system further away from the predefined measurement basis (e.g., the $X$ basis), thereby introducing greater measurement uncertainty. Another important observation is that $P_{e}$ is not reduced with the increase of sensing qubits $N$. Rather, more sensing qubits incurs more accumulation of crosstalk, worsening the decoding performance. This becomes more pronounced for high $J$'s. 
Therefore, considering both the decoding performance and the limited qubit resource, we adopt $N=2$ in the subsequent simulations.

\begin{figure*}[htbp]
    \centering
    \subfloat[$J=0.01$]{
        \includegraphics[width=0.3\textwidth]{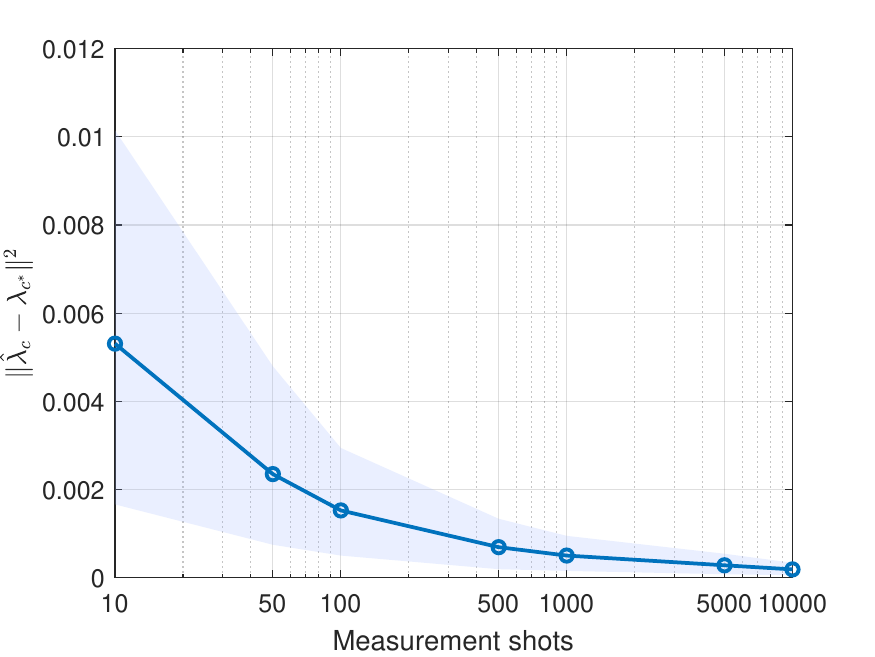}}
    \hfill
    \subfloat[$J=0.1$]{
        \includegraphics[width=0.3\textwidth]{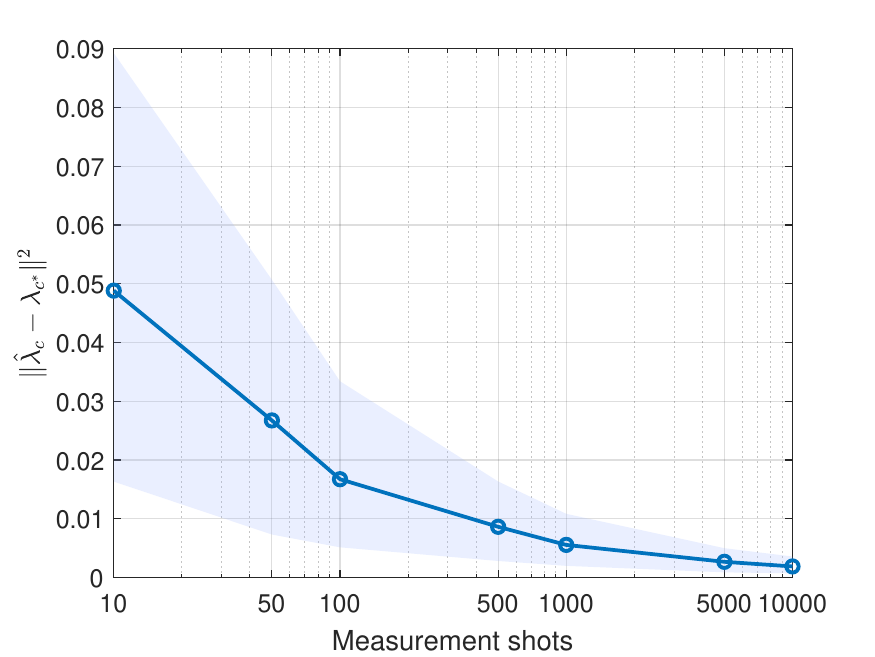}}
    \hfill
    \subfloat[$J=0.5$]{
        \includegraphics[width=0.3\textwidth]{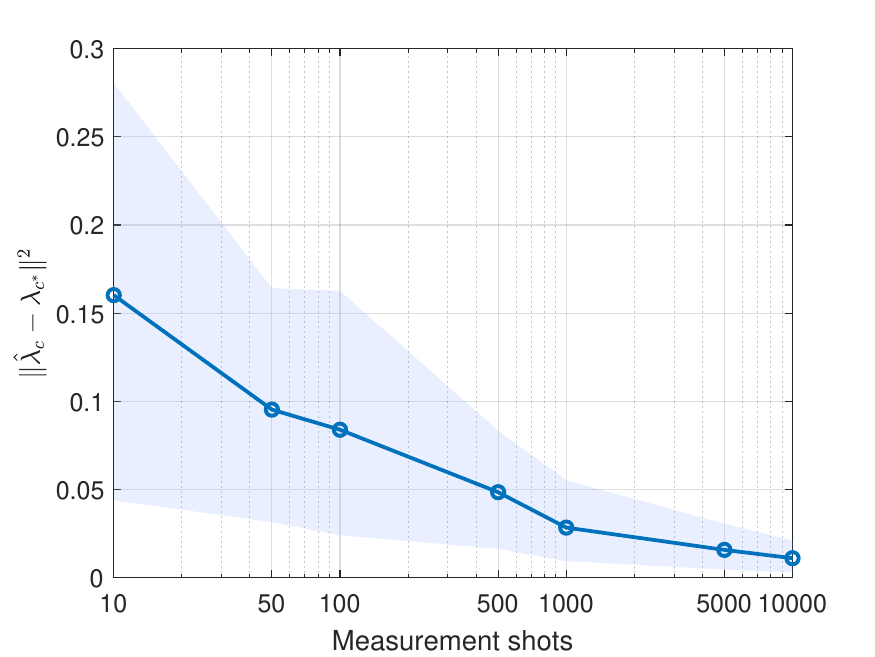}}
        
    \caption{Distribution of eigenvalue distance versus measurement shots $F$ for different $J$'s.}
    \label{fig:dist_vs_shots}
\end{figure*}

Fig.~\ref{fig:dist_vs_shots} illustrates the distribution of eigenvalue distance versus the number of measurement shots $F$. Fig.~\ref{fig:dist_vs_shots}~(a), (b), and (c) correspond to the cases of $J=0.01$, $J=0.1$, and $J=0.5$, respectively.
The blue shaded region represents the $10\%\text{--}90\%$ percentile range of the eigenvalue distance 
$\left\| \hat{\boldsymbol{\lambda}}_{c} - \boldsymbol{\lambda}_{c^*} \right\|^2$, 
where $\hat{\boldsymbol{\lambda}}_{c}$ denotes the estimated eigenvalues evaluated for a given number of measurement shots, and 
$\boldsymbol{\lambda}_{c^*}$ denotes the reference eigenvalues stored in the offline database 
for the optimal signal combination $c^*$ obtained via~(\ref{eq:lambda ML}). It can be observed that this shaded region, which reflects the measurement uncertainty, shrinks as the number of measurement shots $F$ increases. This behavior is consistent with the quantum noise model described in (\ref{eq:var obs}). Furthermore, by comparing Fig.~\ref{fig:dist_vs_shots}~(a), (b), and (c), we observe that achieving the same level of measurement uncertainty requires a larger number of measurement shots when the coupling strength $J$ increases. The underlying reason for this phenomenon has been illustrated in Fig.~\ref{fig:hyperparameter_selection}.

Considering that a large number of measurement shots corresponds to a longer symbol duration, which reduces the achievable symbol rate, we select $F = 10^2$ for the case of coupling strength $J = 0.01$ as a trade-off between performance and transmission efficiency.

\begin{figure}[!t]
\centering
  \includegraphics[width=0.38\textwidth]{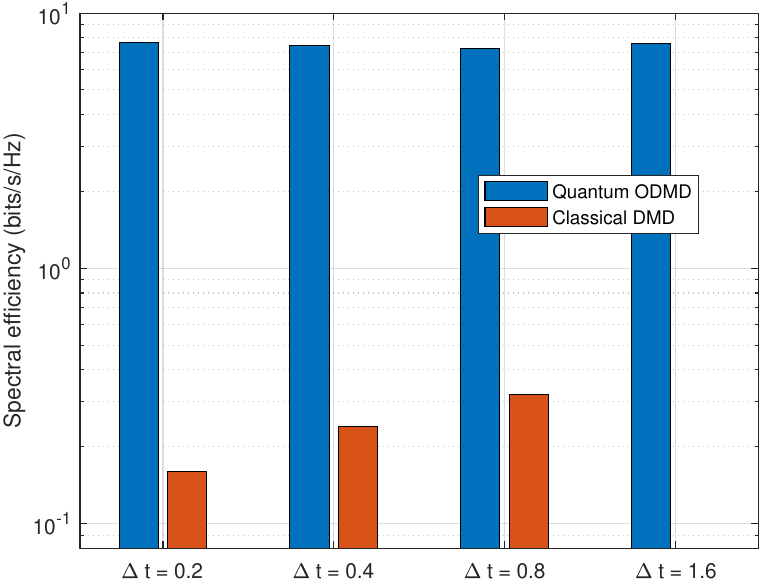}
   \caption{Compare classical and quantum DMD methods.}\label{fig:compDMD}
\end{figure}

\subsubsection{Quantum Advantage}
Fig.~\ref{fig:compDMD} compares the spectral efficiency of the proposed quantum receiver (Quantum ODMD) and classical DMD under different time sampling gaps $\Delta t$ when $K=2$. Classical DMD is applied directly to the received signal of a conventional receiver without quantum processing, and the maximum number of time-sample modes is fixed at $S = 3$. The proposed Quantum ODMD demonstrates strong robustness to variations in the time sampling gap. In contrast, Classical DMD is highly sensitive to the choice of $\Delta t$, as its performance depends critically on carefully designed sampling to accurately capture the underlying signal dynamics.

The higher spectral efficiency of Quantum ODMD arises from the fundamentally different construction of the relation matrix. With the aid of the quantum processor, the least-squares (LS) relation matrix is defined as
\[
A = \left\langle \phi_0 \Big|  \exp\Big(-i \int_{s\Delta t}^{(s+1)\Delta t} H[x(t)]\, dt \Big) \Big| \phi_0 \right\rangle,
\]
which coherently captures all components of the transmitted signal 
\[
x(t) = \sum_k \alpha_k \cos(\omega t + \phi_k).
\] 
In contrast, Classical DMD approximates the relation matrix as 
\[
A_{\text{DMD}} \approx \frac{x((s+1)\Delta t)}{x(s\Delta t)},
\] 
which can lose crucial information, especially when multiple users or phase differences are present.

\begin{figure*}[htbp]
    \centering
        \subfloat[K=1]{\includegraphics[width=0.3\textwidth]{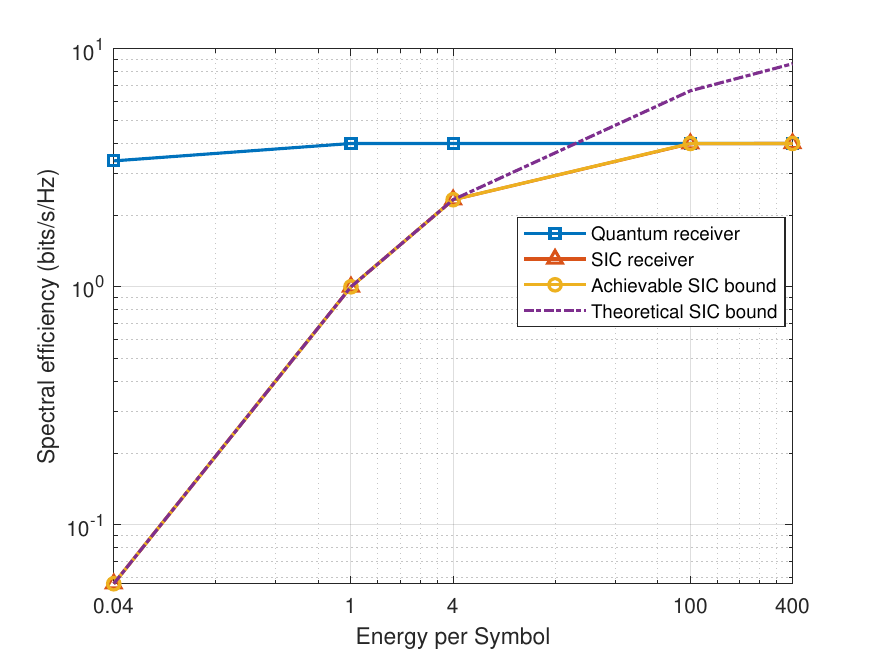}}
    \hfill
    \subfloat[K=2]{
        \includegraphics[width=0.3\textwidth]{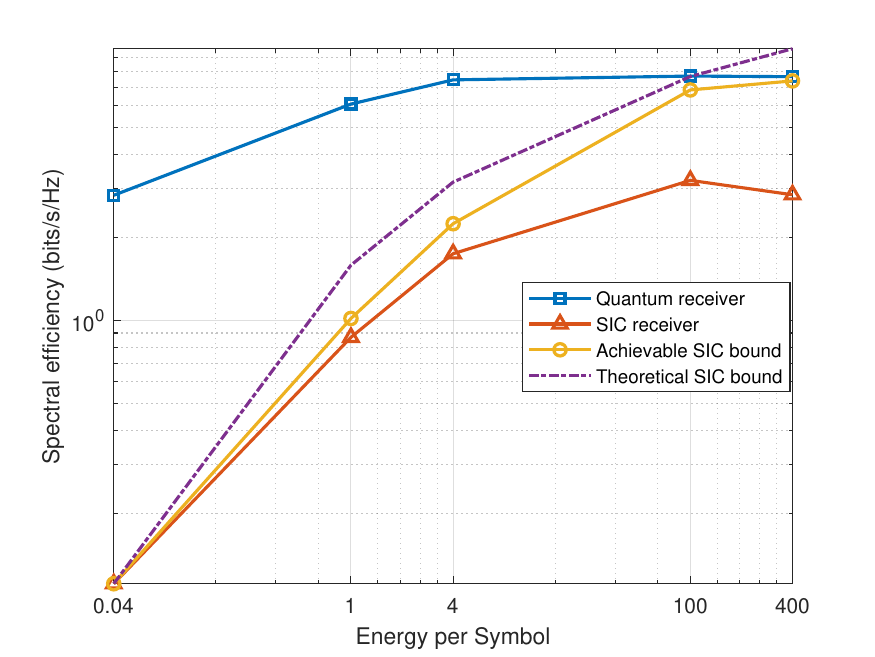}}
    \hfill
    \subfloat[K=3]{
        \includegraphics[width=0.3\textwidth]{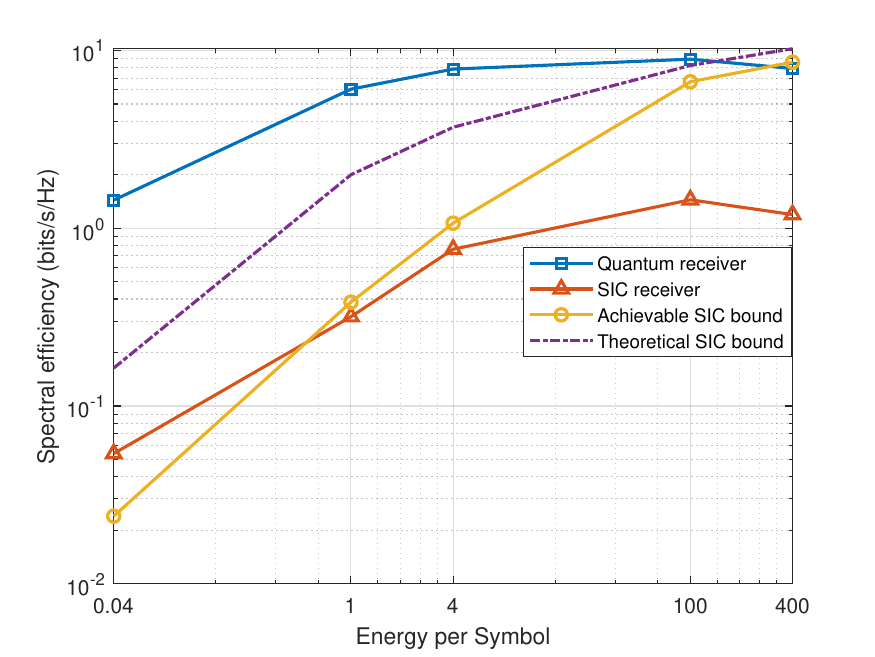}}
        
\caption{Quantum advantage over classical methods for different symbol amplitudes.}

    \label{fig:varAmp-quantum gain}
\end{figure*}
Fig.~\ref{fig:varAmp-quantum gain} illustrates the quantum gain achieved by the proposed quantum receiver compared with classical methods, i.e., the achievable SIC bound (the JMLD receiver) and the SIC receiver, under different energy budgets per transmitted symbol, $\mathrm{En} = 2\pi\cdot\{0.04, 1, 4, 100, 400\}$. The energy budget is defined as
\[
\mathrm{En} = P \cdot |x|^2 \cdot T = P \cdot |x|^2 \cdot F \cdot T_p.
\]

It is observed that the proposed quantum receiver can outperform the theoretical SIC bound when the transmit energy budget is low, specifically under multiple access case $K>1$, for $\mathrm{En} < 2\pi\cdot100$, i.e., $\text{SNR} = \frac{P \cdot |x|^2}{n} < 0\,\mathrm{dB}$. It also surpasses the best achievable SIC limit in the low-SNR regime ($\mathrm{En} < 2\pi\cdot 400$, i.e., $\text{SNR} < 6\,\mathrm{dB}$). However, the quantum advantage in spectral efficiency gradually diminishes as the SIC performance improves in the high-SNR regime. These results highlight the capability of the proposed real-time quantum receiver in extracting weak signals more effectively than classical methods.

This performance gain arises from the fact that quantum measurements of observables encode signal-induced system dynamics in a manner that is inherently immune to classical noise accumulation. In the weak-signal regime, classical receivers operate under severely limited SNR, leading to significant performance degradation. In contrast, real-time snapshot measurements from the quantum system preserve the underlying signal-dependent dynamics with higher fidelity. This enables more accurate reconstruction and detection through the ODMD-based processing framework.

Moreover, it can be seen that the Achievable SIC bound (achieved by the JMLD method) and the SIC receiver exhibit almost identical performance in a single-access network ($K=1$). However, in a multiple-access network ($K=2$ and $K=3$), the Achievable SIC bound outperforms the SIC receiver. This is because, without optimizing the transmit power of multiple users sharing the communication channel, SIC suffers from additional interference among users. In contrast, directly decoding the combination of multiple users' signals, as in the Achievable SIC bound, achieves better performance. Notably, with an appropriate power allocation strategy, the performance of the SIC receiver can be made almost identical to the Achievable SIC bound.

\begin{figure*}[htbp]
    \centering
        \subfloat[K=1]{
        \includegraphics[width=0.3\textwidth]{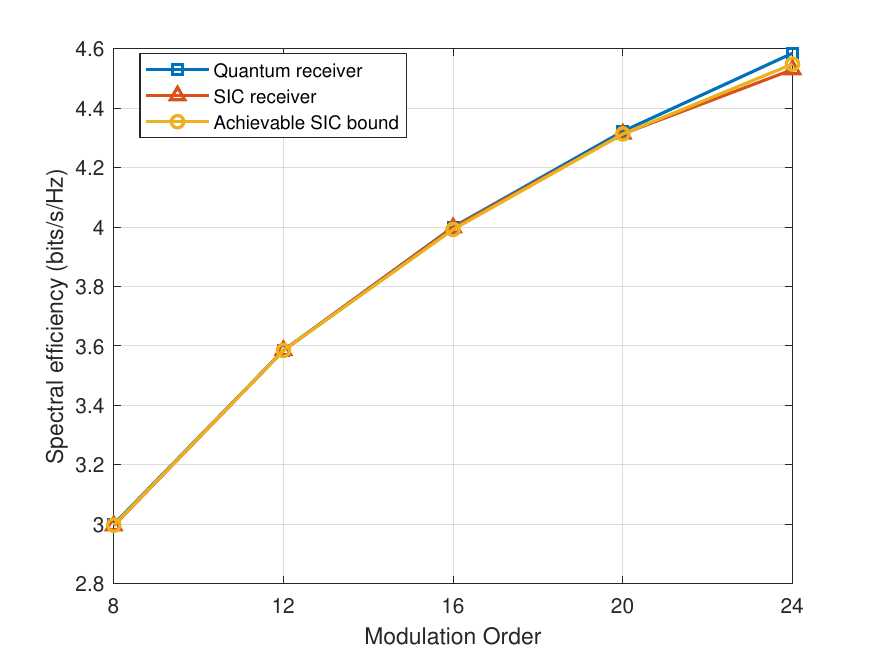}}
    \hfill
    \subfloat[K=2]{
        \includegraphics[width=0.3\textwidth]{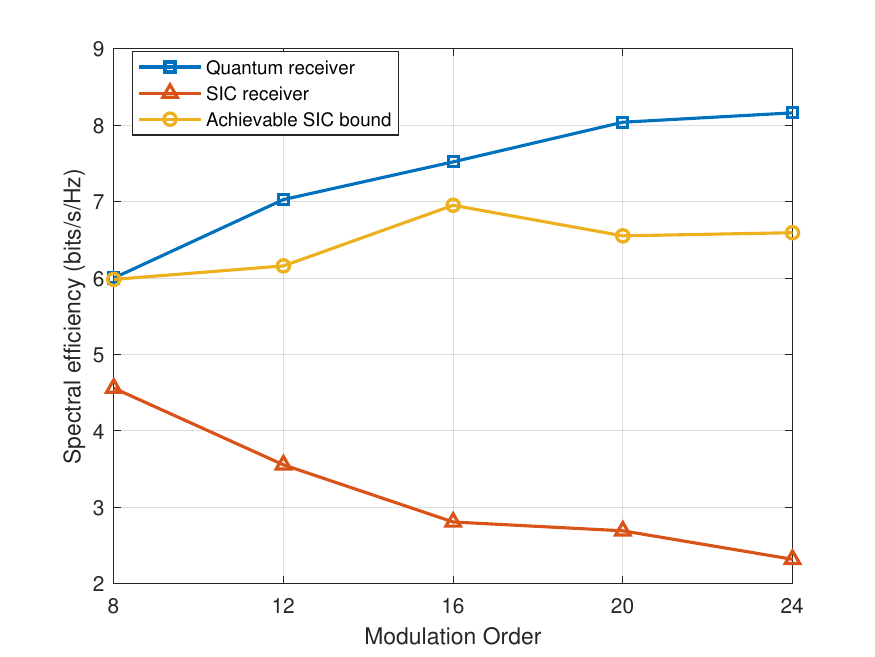}}
    \hfill
    \subfloat[K=3]{
        \includegraphics[width=0.3\textwidth]{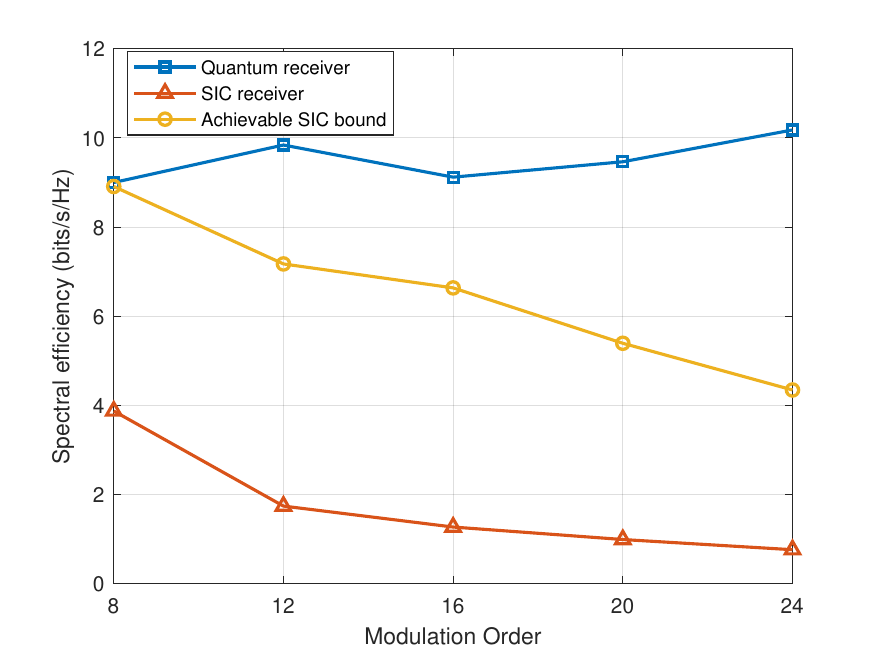}}
        
\caption{Spectral efficiency for different receivers under 4-phase and varying-amplitude modulation.}
    \label{fig:varBit-phase-4}
\end{figure*}

Fig.~\ref{fig:varBit-phase-4} illustrates the spectral efficiency of the proposed quantum receiver and two classical SIC decoding methods for varying modulation orders $M = M_{\alpha} \times 4$. It is observed that the spectral efficiency of the proposed quantum receiver increases with the modulation order. In contrast, when $K = 3$, the spectral efficiency of both the Achievable SIC bound and the SIC receiver decreases as the modulation order increases. This degradation in classical methods is attributed to the reduced minimum distance between signal constellation points at higher modulation orders, which makes reliable symbol discrimination more challenging. In comparison, the proposed quantum receiver can effectively capture finer distinctions between signal codes by leveraging the LS relationship constructed from successive time-snapshot quantum observables. This provides a more expressive representation than the linear models used in the Achievable SIC bound and SIC receiver. As a result, the proposed quantum receiver demonstrates strong potential to further improve spectral efficiency when combined with high-order, carefully designed modulation schemes.

\subsection{Discussion}
Intuitively, the aforementioned quantum advantage stems from two reasons. The first reason lies in the ability to perform temporally correlated measurements on the same underlying received field through sequential sensor operations. Specifically, the sensing qubit can be reset to its initial state before each measurement, enabling multiple independent shots of the same signal within a single symbol duration. This reset mechanism ensures that the measurement noise across different shots is independent and identically distributed, so that averaging effectively reduces the measurement noise without requiring additional channel uses. In contrast, classical receivers with AWGN noises obtain only a single noisy observation per transmitted symbol, and performance improvement typically requires additional channel uses or increased bandwidth.

The second reason is its robustness to measurement noise. The proposed ODMD method suppresses quantum measurement noise by filtering eigen-components through a flexible SVD-based thresholding strategy, thereby achieving a balance between noise suppression and signal preservation. Moreover, ODMD can be naturally extended beyond Hadamard test-based estimation to classical-shadow-style measurements using randomized Pauli observables \cite{shen2026efficient}. This extension reduces the number of required measurement shots while still preserving accurate reconstruction of the signal-induced quantum dynamics.

While this paper assumes $F$ repetitions of baseband signals within a symbol duration $T$, this assumption can be eliminated by using $F$ copies of $S$-mode sensing qubits. For example, in the earlier setup of $F=100$ signal repetitions (i.e., measurement shots) and $S\cdot N = 6$ sensing qubits, we could instead use $F \cdot S \cdot N = 600$ sensing qubits to complete signal detection within one signal cycle. Given that contemporary quantum systems can readily support hundreds of qubits, this approach is entirely practical.  Thus, by exchanging signal-level time redundancy for qubit-level hardware redundancy, the proposed quantum receiver becomes adaptable to a wide range of wireless symbol designs.

\section{Conclusion}\label{sec:conclusion}
This paper reported quantum advantage over the classical SIC technique in multiple access networks in low SNR regimes. This is realized by a novel quantum receiver architecture consisting of front-end quantum sensing and back-end quantum signal processing units. Without using complex quantum circuits or fragile quantum resources like entanglement, the quantum receiver manages to excel in spectral efficiency and signal processing speed. Notably, specifically under multiple access case, the receiver beat the theoretical multiple-access channel capacity limit in ultra-low SNRs (i.e., SNR $<$ 0 dB) and beat the best achievable SIC limit in low SNRs (i.e., SNR $<$ 6 dB), despite that the quantum advantage in spectral efficiency diminishes when the SIC performance improves in high SNRs. We also observe that the quantum receiver can detect data symbols before the end of symbol duration because the receiver exploits the temporal correlation between samplings, resulting in much faster detection speed over classical receivers. 

In conclusion, the proposed quantum receiver demonstrates great promises in highly challenging and contested environments. Future work includes (1) the search among different qubit species (thus different Hamiltonian) for the best candidate for the quantum receiver and (2) the demonstration of potential quantum advantage over other multi-access technologies.

\bibliographystyle{IEEEtran}
\bibliography{ref}

\end{document}